\documentclass[preprint,12pt,a4paper]{elsarticle}

\usepackage{amssymb}
\usepackage{amsmath}
\usepackage[hidelinks,hypertexnames=false]{hyperref}
\usepackage{graphicx}
\usepackage{tikz}
\usepackage{listings}
\usepackage{xcolor}
\usepackage[normalem]{ulem}
\usepackage{booktabs}
\usepackage{makecell}
\usepackage{tabularx}
\usepackage{array}
\usepackage{url}
\usepackage{float}

\usetikzlibrary{arrows.meta,positioning,fit,backgrounds}
\newcommand{\xxyyscript}{{\small\texttt{examples/\allowbreak xxyy\_\allowbreak hamiltonian\_\allowbreak usage\_\allowbreak example.py}}}
\newcommand{\clockscript}{{\small\texttt{examples/\allowbreak quantum\_\allowbreak clock\_\allowbreak model\_\allowbreak usage\_\allowbreak example.py}}}

\journal{SoftwareX}

\begin{document}

\begin{frontmatter}

\title{\texttt{autotn}: Automata-inspired construction of tensor-network operators from symbolic local rules}

\author[deusto,tecnalia]{\texorpdfstring{\href{https://orcid.org/0009-0003-4148-8874}{Aitor Morais}\corref{cor1}}{Aitor Morais}}
\ead{aitor.morais@deusto.es}

\author[upvehu]{\texorpdfstring{\href{https://orcid.org/0000-0002-3950-1668}{Izaskun Oregi}}{Izaskun Oregi}}

\author[deusto]{\texorpdfstring{\href{https://orcid.org/0000-0002-3068-6248}{Iker Pastor}}{Iker Pastor}}

\author[tecnalia]{\texorpdfstring{\href{https://orcid.org/0000-0001-7863-9910}{Eneko Osaba}}{Eneko Osaba}}

\cortext[cor1]{Corresponding author.}

\address[deusto]{University of Deusto (D4K)}
\address[upvehu]{Department of Applied Mathematics, University of the Basque Country (UPV/EHU)}
\address[tecnalia]{Tecnalia, Basque Research and Technology Alliance (BRTA)}

\begin{abstract}
Matrix product operators (MPOs) are a standard representation for structured operators in tensor-network simulations and quantum computing applications. However, converting symbolic descriptions of Hamiltonians and cost functions into MPO representations often requires substantial manual implementation, making the process error-prone and difficult to reproduce. This paper presents \texttt{autotn}, an open-source Python software tool that converts symbolic operator rules into MPO representations. Users define local operators and interaction rules in a high-level format, and the software generates the corresponding MPO along with validation utilities. In doing so, the software follows an automata-inspired approach that identifies repeated patterns in the symbolic description of the operator and reuses them to build a compact tensor-network representation, avoiding the need to construct each tensor by hand. The current implementation of \texttt{autotn} supports both diagonal operators, commonly encountered in combinatorial optimization problems, and general local matrix operators used in quantum many-body physics. The software is demonstrated on Max-Cut cost operators, long-range $XX+YY+Z$ spin Hamiltonians, and quantum clock models. By providing a reproducible workflow from symbolic model specification to validated MPO construction, \texttt{autotn} reduces the engineering effort required to develop, verify, and compare MPO constructions.

\end{abstract}


\end{frontmatter}

\section*{Required Metadata}

\section*{Current code version}

\begin{table}[H]
\centering
\begin{tabularx}{\linewidth}{|l|>{\raggedright\arraybackslash}X|>{\raggedright\arraybackslash}X|}
\hline
\textbf{Nr.} & \textbf{Code metadata description} & \textbf{Current information} \\
\hline
C1 & Current code version & 1.0.0 \\
\hline
C2 & Permanent GitHub link to code/repository used for this code version & \href{https://github.com/aitormorais/tngenerator/tree/v1.0.0}{GitHub tag v1.0.0}. \\
\hline
C3 & Legal Code License & MIT License. \\
\hline
C4 & Code versioning system used & Git, with source code hosted on GitHub at \href{https://github.com/aitormorais/tngenerator}{aitormorais/tngenerator}. \\
\hline
C5 & Software code languages, tools, and services used & Python; NumPy; pytest for tests; Matplotlib for optional benchmark plots; setuptools for packaging. \\
\hline
C6 & Compilation requirements, operating environments \& dependencies & Python $\geq$ 3.10; NumPy. Test dependency: pytest. Optional benchmark plotting dependency: Matplotlib. \\
\hline
C7 & If available Link to developer documentation/manual & Developer documentation is provided in the repository \href{https://github.com/aitormorais/tngenerator/blob/v1.0.0/README.md}{README.md}. \\
\hline
C8 & Support email for questions & \href{mailto:aitor.morais@deusto.es}{aitor.morais@deusto.es}. \\
\hline
\end{tabularx}
\caption{Code metadata.}
\end{table}

\section{Motivation and significance}

Many applications in combinatorial optimization, quantum computing, and quantum many-body physics are defined through sums of local interactions. In the weighted Max-Cut problem, for example, each edge contributes a term involving the variables associated with its two incident vertices. Similarly, quantum many-body Hamiltonians commonly combine local fields and interactions acting on one, two, or several sites \cite{lucas2014ising,schollwoeck2011dmrg}. In these domains, operators are typically specified using symbolic expressions composed of local operator terms and associated coefficients. Such representations are concise, intuitive, and closely aligned with the mathematical formulation of the underlying model.

A generic operator can be written as
\begin{equation}
   \hat{H}
=
\sum_{\alpha=1}^{M}
c_{\alpha}
\bigotimes_{i=1}^{N}
\hat{O}_{i}^{(\alpha)}, \label{eq1:general_hamiltonian}
\end{equation}
where $N$ denotes the number of sites, $M$ is the number of interaction terms, and $c_\alpha$ is the coefficient associated with term $\alpha$. For each site $i$, the local object $\hat{O}_{i}^{(\alpha)}$ represents either a local operator acting on site \(i\) or the identity when that site does not participate in the interaction. This notation is compact and closely reflects the mathematical structure of the underlying model, making it convenient for defining and interpreting both
optimization objectives and physical Hamiltonians \cite{crosswhite2008finite,hubig2017generic}.

However, symbolic expressions alone are not sufficient for numerical calculations. Most computational frameworks require an explicit numerical representation of the operator. Constructing the corresponding dense matrix quickly becomes impractical because its size grows exponentially with the number of sites. Consequently, even operators defined by a relatively small set of symbolic terms can become prohibitively expensive to store and manipulate in their full matrix form. This motivates the use of numerical representations that preserve the local structure of the model without constructing the full operator \cite{orus2014tnreview}.

To address this limitation, tensor-network (TN) representations provide compressed numerical descriptions that exploit the locality and repeated structure present in many operators. Among them, matrix product operators (MPOs, \cite{pirvu2010matrix}) have become a standard representation for Hamiltonians, optimization cost functions, time-evolution operators, and other structured operators \cite{pirvu2010matrix,schollwoeck2011dmrg, haegeman2017diagonalizing,lucas2014ising}. By decomposing a large operator into a sequence of local tensors connected by auxiliary bonds, MPOs can represent many physically relevant operators using substantially fewer parameters than a dense matrix representation.

Despite their widespread use, operators are seldom expressed directly as MPOs. Instead, researchers commonly begin with symbolic descriptions such as Equation \eqref{eq1:general_hamiltonian} and subsequently translate them into numerical MPO representations. This issue has motivated several approaches for systematic MPO construction. Finite-state-machine and automata-based descriptions provide a way to encode families of operator strings into compact MPO structures~\cite{crosswhite2008finite}. Other works construct generic MPOs through MPO arithmetic followed by compression, showing that efficient representations can be obtained for broad classes of operators~\cite{hubig2017generic}. Additional algorithmic schemes exploit graph-based or symmetry-aware constructions to generate MPOs automatically from structured symbolic input~\cite{paeckel2017automated,ren2020automatic}. These developments highlight that MPO construction is not merely an implementation detail; it is the step that connects a mathematical model with the numerical representation required for computations.

The need for MPO construction arises within a mature ecosystem of TN and quantum-simulation software. ITensor provides a comprehensive TN library with mature MPO functionality~\cite{fishman2022itensor}. TeNPy supports Python simulations of strongly correlated quantum systems~\cite{hauschild2018tenpy}. Quimb combines TN algorithms with quantum-information and many-body utilities~\cite{gray2018quimb}, while OpenFermion focuses on symbolic fermionic and qubit-operator workflows for quantum simulation~\cite{mcclean2020openfermion}. Although existing TN libraries provide robust simulation capabilities, the step that converts symbolic model definitions into validated MPO representations is often left to user-specific implementations. This creates an engineering gap between the mathematical specification of an operator and its reliable numerical realization. In this work we present \texttt{autotn} a software designed to bridge this gap by providing a reproducible and automated MPO-construction workflow.
 
The exact MPO representations are derived from symbolic sums of local operator strings. The package provides a reproducible pipeline that transforms compact symbolic model specifications into MPOs, reducing the need for manual tensor assembly. Its design is inspired by the finite-automata approach of Crosswhite and Bacon \cite{crosswhite2008finite}, where automata-like state caching can be used to organize matrix product algorithms. The current implementation does not expose a formal finite-automaton language. Instead, it uses an automata-inspired construction in which groups of terms with common prefixes and suffixes define the virtual states of the TN.

Rather than replacing existing TN and symbolic-operator frameworks, \texttt{autotn} complements them by focusing on the automated generation and validation of MPO representations. Its intended use is reproducibility-oriented prototyping: users define local operators and symbolic interaction rules, while the software constructs the corresponding tensor network, selects an internal split point, and records structural metadata. The resulting MPOs can then be inspected, validated against dense references for small systems, or integrated into larger simulation workflows.

The resulting workflow supports the development and evaluation of MPO encodings by integrating construction, structural diagnostics, and dense-reference validation within a common framework. Tasks that often require problem-specific implementations -- such as split-point selection, tensor-shape verification, and operator validation -- can therefore be performed using the same software infrastructure. As a result, MPO representations can be developed, inspected, validated, and compared more systematically before being incorporated into larger simulation or optimization workflows.

\section{Software description}
This section describes how \texttt{autotn} transforms symbolic operators into MPO representations. We first describe the main software components and then follow the construction flow from model-specific input data to symbolic operator strings, center selection, backend compilation, and validation metadata. The goal is to clarify how the different modules interact, highlight the main design choices of the package, and focus on the functions that are most relevant to the construction process.

\subsection{Workflow}

\begin{figure}[!htbp]
\centering
\resizebox{\linewidth}{!}{%
\begin{tikzpicture}[
    font=\small,
    block/.style={draw, align=center, minimum width=3.6cm, minimum height=0.95cm, inner sep=4pt},
    process/.style={draw, align=center, minimum width=4.2cm, minimum height=0.95cm, inner sep=4pt},
    backend/.style={draw, align=center, minimum width=4.0cm, minimum height=0.95cm, inner sep=4pt},
    group/.style={draw, dashed, rounded corners, inner sep=8pt},
    arrow/.style={-{Latex[length=2.2mm,width=1.6mm]}, line width=0.5pt}
]
\node[block] (params) {Model parameters\\{\footnotesize user data}};
\node[block, right=0.8cm of params] (store) {Local operator store\\{\footnotesize returned by \texttt{store\_matrix()}}};
\node[block, right=0.8cm of store] (terms) {Symbolic terms\\{\footnotesize returned by \texttt{symbolic\_terms()}}};

\node[process, below=1.1cm of store] (model) {Concrete model object\\{\footnotesize \texttt{autotn.models}}};

\node[process, below=1.1cm of model] (generator) {\texttt{TNGenerator}\\{\footnotesize \texttt{autotn.generator}}};
\node[process, below=0.8cm of generator] (frontend) {Symbolic frontend\\{\footnotesize \texttt{autotn.symbolic}}};
\node[process, below=0.8cm of frontend] (center) {Prefix/suffix grouping\\and center selection\\{\footnotesize \texttt{autotn.grouping}, \texttt{autotn.center}}};

\node[backend, below left=1.1cm and 0.6cm of center] (diagonal) {Diagonal backend\\{\footnotesize \texttt{autotn.diagonal}}};
\node[backend, below right=1.1cm and 0.6cm of center] (operator) {Dense-operator backend\\{\footnotesize \texttt{autotn.operator}}};

\node[process, below=2.8cm of center] (output) {MPO / TN arrays\\and compilation metadata\\{\footnotesize \texttt{list[np.ndarray]}, \texttt{CompilationInfo}}};

\draw[arrow] (params.south) -- (model.north west);
\draw[arrow] (store.south) -- (model.north);
\draw[arrow] (terms.south) -- (model.north east);
\draw[arrow] (model) -- (generator);
\draw[arrow] (generator) -- (frontend);
\draw[arrow] (frontend) -- (center);
\draw[arrow] (center) -- (diagonal);
\draw[arrow] (center) -- (operator);
\draw[arrow] (diagonal.south) -- (output.north west);
\draw[arrow] (operator.south) -- (output.north east);

\begin{scope}[on background layer]
\node[group, fit=(params)(store)(terms), label={[font=\small]above:User input}] {};
\node[group, fit=(model), label={[font=\small]left:Model layer}] {};
\node[group, fit=(generator)(frontend)(center)(diagonal)(operator), label={[font=\small]left:\texttt{autotn} engine}] {};
\node[group, fit=(output), label={[font=\small]below:MPO output}] {};
\end{scope}
\end{tikzpicture}%
}
\caption{Software architecture used by \texttt{autotn}. The user provides problem parameters and the symbolic operators. The model layer gathers input information for the rest of the package. The \texttt{autotn} engine processes this symbolic description, selects a construction route, and produces an MPO representation together with compilation metadata.}
\label{fig:software-architecture}
\end{figure}
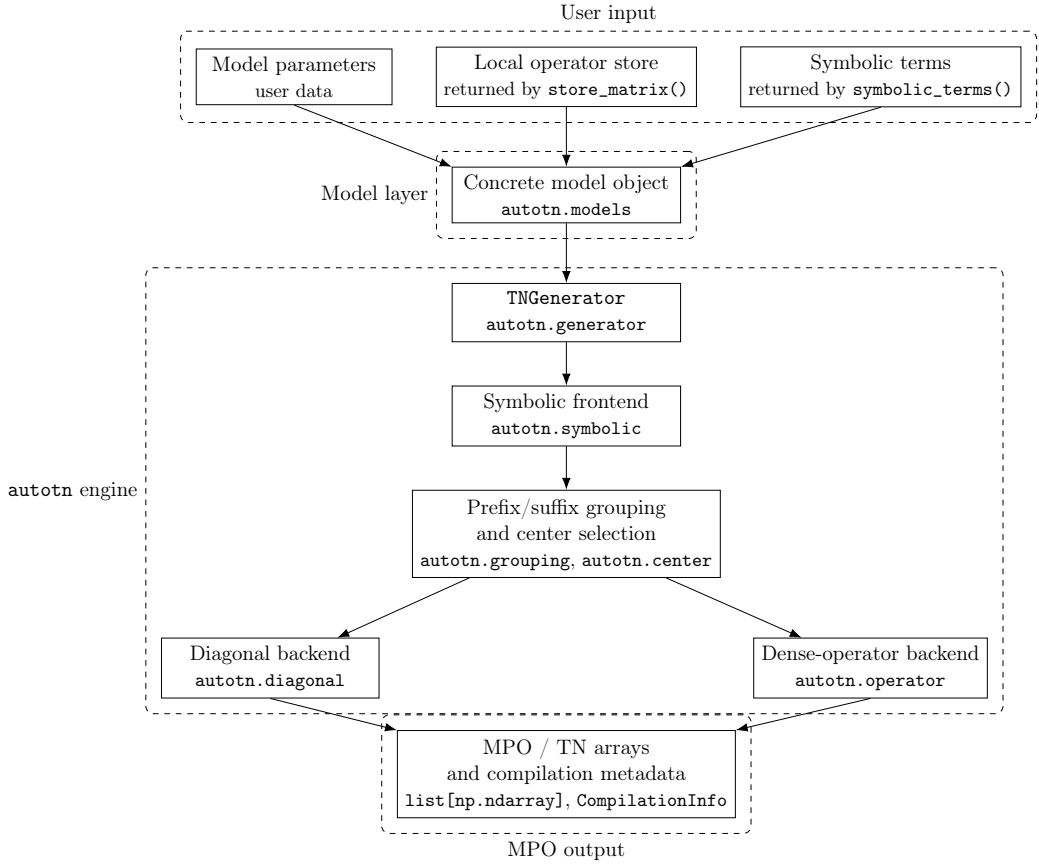

The architecture of \texttt{autotn} is organized into several stages transforming user input into an MPO. As shown in figure \ref{fig:software-architecture}, the workflow starts from the problem-specific model parameters provided by the user. These parameters are encapsulated in a model object implemented in \texttt{autotn.models}. The model object provides \texttt{store\_matrix()} which contains operators in matrix or vector forms, and \texttt{symbolic\_terms()} which contains all the symbolic operator terms.

For model-based construction, users call the \texttt{TNGenerator} class.
Given a model object, \texttt{TNGenerator} retrieves
the local operator store, passes the model's \texttt{symbolic\_terms()} method as
the symbolic term generator, selects any model-provided default center strategy
when appropriate, and delegates the remaining compilation workflow to
\texttt{autotn.symbolic}. The symbolic frontend normalizes the symbolic
description, infers the number of sites when it is not provided explicitly, and
determines whether the local store contains diagonal vectors or dense
local-operator matrices.

The automata-inspired part of the workflow appears in the prefix/suffix grouping
and center-selection stage. Symbolic operator strings are grouped according to
shared prefixes and suffixes, and these shared partial strings define the virtual
states used to build the TN representation. The same prefix/suffix
structure is also used to compare candidate split points before constructing the
final tensors.

The construction is then dispatched to one of two backend modules. Diagonal
local-vector descriptions are compiled by \texttt{autotn.diagonal}, while dense
local-operator descriptions are compiled by \texttt{autotn.operator}. In both
cases, the selected backend produces the final MPO/TN arrays together with
compilation metadata describing the selected construction.

\subsection{Software functionalities}

\texttt{autotn} is organized around a compact set of construction, inspection,
and validation tasks for TN operators. The current repository focuses
on converting symbolic local-operator rules into concrete NumPy tensor arrays,
while exposing enough diagnostics to compare models, split choices, and small
dense-reference checks. The repository currently provides:

\begin{itemize}
    \item construction of diagonal TN from symbolic local-vector rules;
    \item construction of dense local-operator TN from symbolic local-matrix rules;
    \item automatic split-point selection using exact shape estimates;
    \item validation utilities for small systems by explicit contraction;
    \item benchmark scripts that summarize tensor sizes, virtual bond dimensions, build times, and dense-validation errors;
    \item a Max-Cut cost-operator example using diagonal \(Z_i Z_j\) terms;
    \item a long-range \(XX+YY+Z\) Hamiltonian example using dense Pauli operators;
    \item a quantum clock Hamiltonian example using generalized Pauli operators for qudits;
    \item unit tests for token parsing, grouping, center selection, tensor construction, Hamiltonian helpers, and the public front-end.
\end{itemize}

\subsection{Automata-inspired TN construction}

Based on \cite{crosswhite2008finite}, the MPO construction implemented in this software is based on the idea that different symbolic operator strings often share common partial structures. Instead of treating every term independently, the software identifies repeated prefixes and suffixes among the symbolic strings and reuses them when constructing the virtual bonds of the MPO. In this sense, the virtual bonds act as cached partial histories of the symbolic operator strings.

Formally, let \(T=\{t_1,\ldots,t_m\}\) be a set of symbolic operator strings, where each string \(t_k=(s_{k,1},\ldots,s_{k,N})\) contains one local token\footnote{In this context, a token denotes the symbolic label of the local object assigned to one site in an operator string. For example, in the string \texttt{ZIZI}, the tokens are \texttt{Z}, \texttt{I}, \texttt{Z}, and \texttt{I}, corresponding to the local operator product \(Z_1 I_2 Z_3 I_4\).} per site. For a candidate split position, \texttt{autotn} groups terms by their left prefixes \((s_{k,1},\ldots,s_{k,i})\) and right suffixes \((s_{k,i+1},\ldots,s_{k,N})\). These equivalence classes become virtual states. 

At site \(i\), the tensor records the allowed transitions between virtual states. A transition is included when at least one term in \(T\) connects the corresponding partial operator strings through its local token at site \(i\).

As a simple example, consider a four-site diagonal operator containing all pairwise $Z$ operations,
\begin{equation}
    \sum_{1\leq i<j\leq 4} Z_i Z_j .
\end{equation}
With the local alphabet \(\mathcal{A}=\{I,Z\}\), the corresponding symbolic
strings are
\begin{equation}
    T=\{ZZII,\;ZIZI,\;ZIIZ,\;IZZI,\;IZIZ,\;IIZZ\}.
\end{equation}
In this example \(N=4\) is the number of sites, \(m=6\) is the number of
symbolic strings, and \(|\mathcal{A}|=2\) is the number of different local
tokens used. For instance, \(t_2=ZIZI\) denotes the local-token tuple
\((Z,I,Z,I)\), corresponding to \(Z_1 I_2 Z_3 I_4\). Several of these strings share common partial patterns. For example, \texttt{ZZII}, \texttt{ZIZI}, and \texttt{ZIIZ} all begin with the same first
token, \texttt{Z}, indicating that the first site already carries a non-identity operator in those terms. Similarly, strings such as \texttt{ZIIZ},
\texttt{IZIZ}, and \texttt{IIZZ} share suffix patterns that end with
\texttt{Z}, indicating that the last site participates in the corresponding
interaction. More generally, common prefixes encode identical partial structures from the left of the chain, while common suffixes encode identical partial structures from the right. \texttt{autotn} reuses these repeated partial structures as virtual states of the MPO instead of treating each symbolic string as an independent term.

This is the sense in which the implementation is automata-inspired. The prefix and suffix classes play a role analogous to states in a finite-state description: they encode the partial symbolic configurations identified before and after a given site. These states are then associated with the virtual bonds of the MPO, allowing repeated symbolic structure to be reused during tensor construction. In \texttt{autotn}, this automata-inspired formulation is implemented through an internal prefix/suffix construction that compiles symbolic operator strings into concrete TN arrays.  The following schematic summarizes the construction:

\begin{enumerate}
    \item Generate symbolic local strings and optional coefficient-vector references.
    \item For each candidate center, estimate the number of prefix/suffix states and tensor entries.
    \item Select the center according to the requested strategy, such as \texttt{default} or \texttt{min\_elements}.
    \item Build only the selected tensor network by resolving symbolic tokens through \texttt{store\_matrix}.
\end{enumerate}

A relevant feature of this workflow is that candidate split points are scored by estimating tensor shapes before construction. The final tensor network is built only for the selected candidate. This is useful in research settings where one wants to inspect memory-related metrics without materializing every possible tensor network.

The current benchmark scripts compare a default center with a \texttt{min\_\allowbreak elements} strategy. This strategy minimizes the total number of tensor entries first and the maximum virtual bond dimension second. It should therefore be interpreted as a size-oriented strategy rather than as a build-time optimization; evaluating candidate centers introduces additional construction overhead.

\section{Illustrative examples}
This section shows how \texttt{autotn} is used in practice. The aim is to move from the internal description of the package to concrete examples that a reader can run, inspect, and adapt. Each example follows the same basic idea: a model is defined through its numerical parameters, the symbolic representation is generated by the corresponding model class, and \texttt{TNGenerator} builds the tensor network together with useful diagnostic information. For small systems, the resulting tensors are also contracted and compared against an independent dense reference, so that the examples illustrate not only how to use the software, but also how its output can be checked. This workflow is intentionally simple, since the goal is to make the construction process transparent before considering larger instances where dense validation is no longer practical.

\subsection{Model/generator workflow}

The executable examples use the same high-level workflow: construct or load model parameters, wrap them in a model object, build the tensor network with \texttt{TNGenerator}, and validate small cases against a dense reference function from \texttt{autotn.references}. The following abbreviated example follows the implemented \(XX+YY+Z\) script in \xxyyscript.

\begin{lstlisting}
import numpy as np
from autotn import TNGenerator
from autotn.operator import dense_operator_from_tn
from autotn.hamiltonians import generate_symmetric_couplings_and_fields
from autotn.models import XXYYZModel
from autotn.references import xxyyz_dense_reference

J, h = generate_symmetric_couplings_and_fields(n_sites=4, seed=123)
model = XXYYZModel(J=J, h=h)
tn, info = TNGenerator(model).build()

contracted = dense_operator_from_tn(tn)
reference = xxyyz_dense_reference(J, h)
assert np.allclose(contracted, reference)

print(info.selected_middle)
print(info.max_bond_dimension)
print(info.total_tensor_elements)
\end{lstlisting}

The example can be read line by line as a compact version of the full workflow.
\textbf{Lines~1--6} import the objects needed to run the construction and validation:
\texttt{numpy} provides the numerical comparison, \texttt{TNGenerator} builds
the tensor network, \texttt{dense\_operator\_from\_tn} contracts it for small
systems, and the remaining imports provide the parameter generator, model
wrapper, and independent dense reference. \textbf{Line~8} creates the numerical
parameters of the problem, namely the couplings \(J\) and fields \(h\), with a
fixed seed so that the example is reproducible. \textbf{Line~9} wraps these arrays in an
\texttt{XXYYZModel}, which exposes the local operator store and symbolic terms
expected by \texttt{autotn}. \textbf{Line~10} is the actual TN construction:
\texttt{TNGenerator(model).build()} returns the tensor list \texttt{tn} and a
metadata object \texttt{info}. \textbf{Lines~12--14} perform the dense small-system
check. The generated tensor network is contracted, an independent dense
Hamiltonian is built directly from \(J\) and \(h\), and both are compared with
\texttt{np.allclose}. Finally, \textbf{lines~16--18} print diagnostic information: the
selected center, the maximum virtual bond dimension, and the total number of
tensor entries.

The returned \texttt{tn} is a list of NumPy arrays. For this dense local-operator
example, each tensor has four axes. The \texttt{info} object records the selected
split point and shape metrics. The validation line compares the contracted
TN operator against an independently constructed dense reference
from \texttt{autotn.\allowbreak references}. This reference implements the Hamiltonian
formula directly from \(J\) and \(h\), rather than reusing the symbolic token
generator.

\subsection{Currently implemented Hamiltonian examples}

Three executable examples are currently included in the repository. Table~\ref{tab:examples}
summarizes their scripts, model wrappers, backends, and dense-reference
functions. They differ in local dimension, backend, and dense-reference formula,
but all use the model/generator interface.

\begin{table}[H]
\centering
\footnotesize
\renewcommand{\arraystretch}{1.12}

\begin{tabular}{
    @{}
    >{\raggedright\arraybackslash}p{0.31\textwidth}
    @{\hspace{4pt}}
    >{\raggedright\arraybackslash}p{0.21\textwidth}
    @{\hspace{4pt}}
    >{\raggedright\arraybackslash}p{0.17\textwidth}
    @{\hspace{4pt}}
    >{\raggedright\arraybackslash}p{0.24\textwidth}
    @{}
}
\toprule
\textbf{Example script}
&
\textbf{Model wrapper}
&
\textbf{Backend}
&
\textbf{Dense reference}
\\
\midrule

\path{examples/maxcut_usage_example.py}
&
\texttt{MaxCutModel}
&
\makecell[l]{Diagonal\\vectors}
&
\makecell[l]{
    \texttt{maxcut\_dense\_}\\
    \texttt{reference(W)}
}
\\[2pt]

\path{examples/xxyy_hamiltonian_usage_example.py}
&
\texttt{XXYYZModel}
&
\makecell[l]{Dense local\\matrices}
&
\makecell[l]{
    \texttt{xxyyz\_dense\_}\\
    \texttt{reference(J,h)}
}
\\[2pt]

\path{examples/quantum_clock_model_usage_example.py}
&
\makecell[l]{
    \texttt{Quantum}\\
    \texttt{ClockModel}
}
&
\makecell[l]{Dense local\\matrices}
&
\makecell[l]{
    \texttt{quantum\_clock\_}\\
    \texttt{dense\_reference}\\
    \texttt{(J,h,q)}
}
\\

\bottomrule
\end{tabular}

\caption{Executable examples included in the repository. The dense reference implementations are independent of the symbolic token generators and are used to validate the resulting tensor networks for small systems.}
\label{tab:examples}
\end{table}

A brief description of each example is provided below.
\begin{enumerate}
    \item \textbf{Max-Cut cost operator.} The example generates a weighted adjacency matrix, wraps it in \texttt{MaxCutModel}, and builds a diagonal TN through \texttt{TNGenerator}. It compares the default center with the model's size-oriented center strategy and checks the optimized tensor network against \texttt{maxcut\_dense\_reference}.
    \item \textbf{Long-range \(XX+YY+Z\) Hamiltonian.} The script \xxyyscript{} generates reproducible \(J\) and \(h\) parameters, builds a dense local-operator tensor network through \texttt{XXYYZModel} and \texttt{TNGenerator}, prints symbolic term metadata, and validates the contracted operator against \texttt{xxyyz\_dense\_reference}.
    \item \textbf{Quantum clock Hamiltonian.} The script \clockscript{} constructs long-range qudit clock Hamiltonians for several \(N\) and \(q\) values. The implementation treats \(q=2\) as a special case in which dagger-paired terms collapse, reports the expected symbolic term count, and validates small dense operators against \texttt{quantum\_clock\_dense\_reference}.
\end{enumerate}

\section{Validation and benchmark results}
\label{sec:validation-benchmark}

\begin{figure}[!htbp]
\centering
\includegraphics[width=\linewidth]{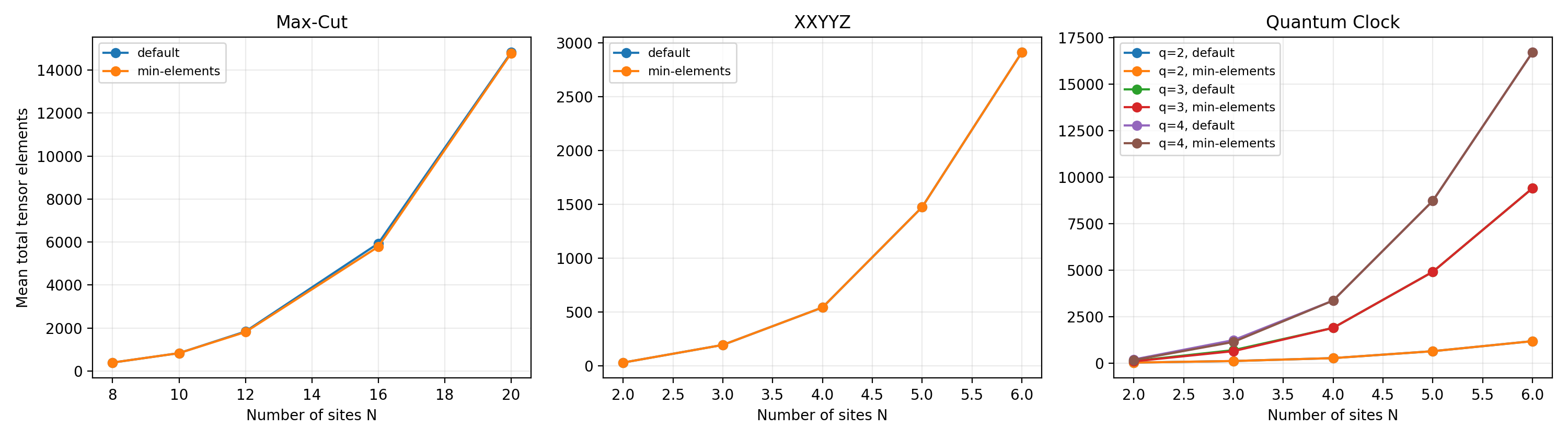}
\caption{Mean total tensor elements for the benchmarked models. The panels separate Max-Cut, XXYYZ, and quantum clock instances. For XXYYZ, the two center strategies overlap because they produce identical tensor sizes in the tested range.}
\label{fig:tensor-elements}
\end{figure}

\begin{figure}[!htbp]
\centering
\includegraphics[width=\linewidth]{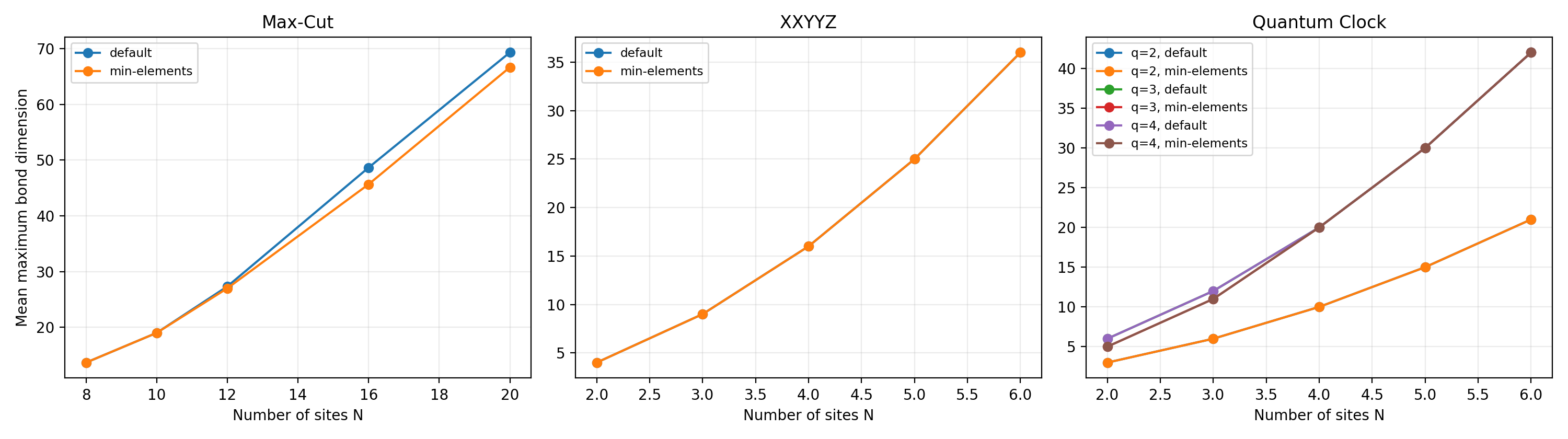}
\caption{Mean maximum virtual bond dimension \(D_{\max}\) for the same build benchmark. The \texttt{min\_elements} strategy is primarily a total-size criterion, so \(D_{\max}\) should be interpreted as a diagnostic rather than as the optimized objective.}
\label{fig:dmax}
\end{figure}

\begin{figure}[!htbp]
\centering
\includegraphics[width=\linewidth]{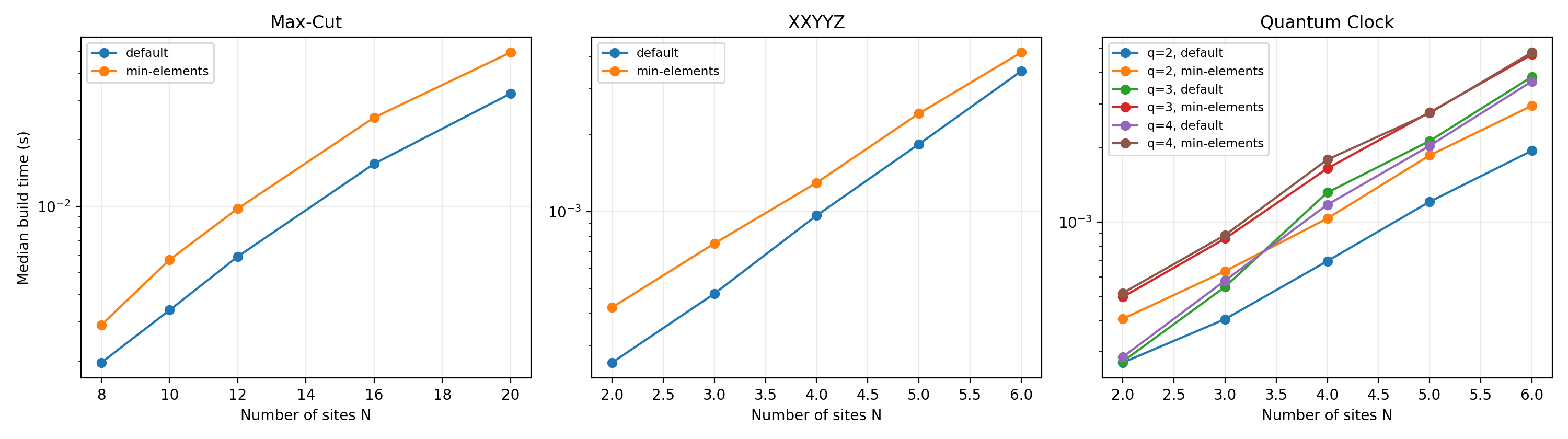}
\caption{Median TN build time. The vertical axis is logarithmic. The \texttt{min\_elements} strategy has additional overhead because it evaluates candidate centers before building the selected tensor network.}
\label{fig:build-time}
\end{figure}

The repository includes two benchmark workflows. The build benchmark measures
construction time and structural TN metrics without dense
validation. The validation benchmark contracts small generated networks and
compares them with dense Hamiltonians built by separate reference functions in
\texttt{autotn.references}. These functions implement the model equations
directly from the numerical parameters, such as \(W\), \(J\), \(h\), and \(q\),
and do not use symbolic token lists or TN construction. This
validation is designed to catch mismatches between symbolic rules and generated
TN arrays, which is precisely the class of implementation error
that often occurs when MPOs or related operator networks are assembled manually.
Each benchmark run also records the symbolic metadata introduced by the
front-end architecture. The fields \texttt{n\_terms} and
\texttt{n\_coefficients} are not performance metrics by themselves: they
document the symbolic workload being compiled and make explicit when several
operator strings share entries in the same coefficient vector. They provide
provenance for the structural measurements reported below, while the visible
benchmark summaries focus on total tensor elements, \(D_{\max}\), construction
time, and dense-validation error.
The benchmark results reported below summarize the build-size and build-time
trade-off in Table~\ref{tab:min-elements-summary} and
Figures~\ref{fig:tensor-elements}--\ref{fig:build-time}, followed by the dense
validation coverage in Table~\ref{tab:validation-summary} and
the error values reported there. These results are intended as reproducibility
and reliability checks for the current implementation, not as
hardware-independent performance claims.

The current build benchmark contains 4500 construction runs, corresponding to
30 repeated timings per tested configuration and center strategy. It compares
the default center with the \texttt{min\_elements} strategy on matched model
configurations. In this comparison, an ``improvement'' means a reduction in the
total number of tensor entries, which is a structural proxy for memory footprint
and TN size. It does not mean that the construction is faster. The
\texttt{min\_elements} strategy evaluates candidate split points before building
the final tensor network, so it can reduce tensor size while still increasing
build time. Table~\ref{tab:min-elements-summary} summarizes this size
size-time trade-off.

\begin{table}[H]
\centering
\resizebox{\linewidth}{!}{%
\begin{tabular}{lrrrrrr}
\toprule
Model & Element ratio & Build-time ratio & Improved & Tied & Worse & Total \\
\midrule
Max-Cut & 0.992 & 1.63 & 6 & 9 & 0 & 15 \\
XXYYZ & 1.000 & 1.41 & 0 & 15 & 0 & 15 \\
Quantum Clock & 0.967 & 1.48 & 12 & 33 & 0 & 45 \\
\bottomrule
\end{tabular}
}
\caption{Summary of the \texttt{min\_elements} strategy relative to the default center. Element ratios are average ratios of mean total tensor elements over matched configurations. Build-time ratios are average ratios of median build time. Improved, tied, and worse count matched configurations for which \texttt{min\_elements} reduced, preserved, or increased the total tensor size. The total differs across model families because quantum clock instances are evaluated for three local dimensions, \(q=2,3,4\), while Max-Cut and XXYYZ have no \(q\) sweep.}
\label{tab:min-elements-summary}
\end{table}

The interpretation is deliberately limited. For Max-Cut, the element ratio is
0.992, indicating a very small average size reduction: six matched
configurations improve and nine are unchanged. For XXYYZ, the ratio is 1.000
and all matched configurations are tied, so the alternative center selection
does not reduce tensor size for this model family in the tested range. The
clearest structural reduction appears for the quantum clock cases, where the
mean element ratio is 0.967 and 12 of 45 matched configurations improve. In all
three model families, however, the median construction time increases by roughly
\(1.4\times\)--\(1.6\times\), because the size-oriented strategy spends extra
time estimating and ranking candidate centers. Thus, \texttt{min\_elements}
should be described as a tensor-size heuristic rather than as a general
performance optimization. Also, because the strategy optimizes total tensor
elements rather than only \(D_{\max}\), occasional increases in maximum bond
dimension are possible when they reduce the total tensor size.
The detailed scaling plots separate these effects: Figure~\ref{fig:tensor-elements}
shows the mean total tensor elements, Figure~\ref{fig:dmax} reports the
corresponding maximum virtual bond dimensions, and Figure~\ref{fig:build-time}
shows the median construction time.

Dense validation addresses a different question: correctness rather than
improvement. It was run for 246 cases across both center strategies. Of these,
192 passed and 54 were skipped because their Hilbert-space dimension was above
the dense-reference limit configured for the corresponding backend. Here this
dimension is the number of basis configurations that the dense reference would
need to represent. For the reported run, the limit was dimension \(4096\) for
diagonal dense-vector validation and dimension \(1024\) for dense
operator-matrix validation. No
dense-validation failures were observed. Table~\ref{tab:validation-summary}
summarizes the current validation coverage. Max-Cut and XXYYZ match exactly
within the implemented floating-point comparison. The complex-valued quantum
clock model reaches a maximum absolute deviation of \(1.95\times 10^{-14}\),
consistent with numerical round-off from dense contraction and complex
arithmetic. This validation does not prove
asymptotic performance or cover arbitrarily large systems, but it directly
verifies that the generated TN operators agree with dense
Hamiltonians constructed directly from \(W\), \(J\), \(h\), and \(q\),
independently of the symbolic token lists.

\begin{table}[H]
\centering
\resizebox{\linewidth}{!}{%
\begin{tabular}{lrrrp{5.0cm}r}
\toprule
Model & Passed & Skipped & Failed & Largest validated cases & Max. abs. error \\
\midrule
Max-Cut & 30 & 0 & 0 & \(N=12\), \(\dim=4096\) & 0.0 \\
XXYYZ & 54 & 0 & 0 & \(N=10\), \(\dim=1024\) & 0.0 \\
Quantum Clock & 108 & 54 & 0 & \(q=2: N=10, \dim=1024\); \(q=3: N=6, \dim=729\); \(q=4: N=5, \dim=1024\) & \(1.95\times 10^{-14}\) \\
\bottomrule
\end{tabular}
}
\caption{Dense-validation summary for the current benchmark data. Skipped cases correspond to configurations above the reported dense-reference Hilbert-space limits: dimension \(4096\) for diagonal dense-vector checks and dimension \(1024\) for dense operator-matrix checks.}
\label{tab:validation-summary}
\end{table}

\section{Impact}

The main contribution of \texttt{autotn} is a reproducible path from symbolic local operator rules to exact TN arrays. It reduces the engineering barrier between model specification and validated TN operator construction. This is useful in research workflows where new Hamiltonian or cost-function encodings must be prototyped, inspected, and compared without repeatedly hand-coding MPO.

\texttt{autotn} separates three concerns that are often mixed: symbolic operator generation, TN construction, and exact small-system validation. This separation helps users identify whether an error comes from the model rule, the tensor construction, or later numerical use of the operator. The structural diagnostics also expose how grouping and center-selection decisions affect \(D_{\max}\), memory-related tensor sizes, and build-time overhead. As a result, the package provides a reusable construction and validation layer that can support later integration with other TN solvers and benchmarking pipelines.

\section{Conclusions and future work}

\texttt{autotn} implements an automata-inspired approach for constructing exact TN representations from symbolic local operator strings. The current repository demonstrates diagonal and dense local-operator compilation, automatic split-point selection, structural diagnostics, and small-system validation against dense references. The design is intentionally compact: users define local arrays and symbolic operator rules, while the software builds and checks the corresponding TN representation.

Table~\ref{tab:feature-scope} summarizes the practical boundary of this first
release. The supported column reflects the part of the workflow that the package
currently owns: turning finite symbolic local-token descriptions into concrete
NumPy TN arrays and checking them on small systems. The unsupported
column is equally important because it separates this construction layer from
tasks that belong to broader TN frameworks or symbolic algebra
systems. In this sense, the current software is best understood as a focused
operator-generation and validation component rather than a complete
simulation environment.

\begin{table}[H]
\centering
\resizebox{\linewidth}{!}{%
\begin{tabular}{>{\raggedright\arraybackslash}p{6.2cm}>{\raggedright\arraybackslash}p{6.2cm}}
\toprule
Supported in the current repository & Not supported in the current repository \\
\midrule
Symbolic local-token descriptions of operator strings & General symbolic algebra or expression simplification \\
Diagonal local-vector TN & DMRG, MPS optimization, or variational solvers \\
Dense local-operator TN & Automatic fermionic sign handling or second-quantized algebra \\
Weighted tokens through coefficient-vector lookup & Large-scale dense validation beyond small Hilbert spaces \\
Automatic center selection by estimated tensor size & Dataset management or full experiment orchestration \\
Dense reference validation for small systems & Replacement of mature TN solver libraries \\
\bottomrule
\end{tabular}
}
\caption{Current feature scope. The software is designed around operator construction and validation, not around solving TN optimization problems.}
\label{tab:feature-scope}
\end{table}

Planned work follows directly from this boundary: improving documentation of the
symbolic front-end, extending benchmark coverage to larger and more diverse
models, and integrating the generated TN with solver workflows.
Additional Hamiltonian examples should be added only once they are implemented, tested,
and included in the validation pipeline.

\section*{Acknowledgements}

Aitor Morais and Iker Pastor acknowledge partial funding for Aitor Morais's doctoral research at the University of Deusto, within the D4K (Deusto for Knowledge) team on applied artificial intelligence and quantum computing technologies.

\section*{CRediT authorship contribution statement}
\textbf{Aitor Morais:} Conceptualization, Investigation, Methodology, Formal analysis, Software, Writing - Original Draft.
\textbf{Eneko Osaba:} Conceptualization, Validation, Investigation, Writing - Review \& Editing, Supervision.
\textbf{Iker Pastor:} Funding Acquisition.
\textbf{Izaskun Oregi:} Conceptualization, Validation, Investigation, Writing - Review \& Editing, Supervision.

\bibliographystyle{elsarticle-num}
\bibliography{softwarex-osp-template}

@article{crosswhite2008finite,
  author = {Crosswhite, Gregory M. and Bacon, Dave},
  title = {Finite automata for caching in matrix product algorithms},
  journal = {Physical Review A},
  volume = {78},
  number = {1},
  pages = {012356},
  year = {2008},
  doi = {10.1103/PhysRevA.78.012356}
}

@article{fishman2022itensor,
  author = {Fishman, Matthew and White, Steven R. and Stoudenmire, E. Miles},
  title = {The {ITensor} Software Library for Tensor Network Calculations},
  journal = {SciPost Physics Codebases},
  pages = {4},
  year = {2022},
  doi = {10.21468/SciPostPhysCodeb.4}
}

@article{hauschild2018tenpy,
  author = {Hauschild, Johannes and Pollmann, Frank},
  title = {Efficient numerical simulations with Tensor Networks: Tensor Network {Python} ({TeNPy})},
  journal = {SciPost Physics Lecture Notes},
  pages = {5},
  year = {2018},
  doi = {10.21468/SciPostPhysLectNotes.5}
}

@article{gray2018quimb,
  author = {Gray, Johnnie},
  title = {quimb: A {Python} package for quantum information and many-body calculations},
  journal = {Journal of Open Source Software},
  volume = {3},
  number = {29},
  pages = {819},
  year = {2018},
  doi = {10.21105/joss.00819}
}

@article{mcclean2020openfermion,
  author = {McClean, Jarrod R. and Rubin, Nicholas C. and Babbush, Ryan and Bacon, Dave and Bushnell, Christa and Boixo, Sergio and McClain, Christopher and Neven, Hartmut},
  title = {OpenFermion: the electronic structure package for quantum computers},
  journal = {Quantum Science and Technology},
  volume = {5},
  number = {3},
  pages = {034014},
  year = {2020},
  doi = {10.1088/2058-9565/ab8ebc}
}

@article{orus2014tnreview,
  title   = {A practical introduction to tensor networks},
  author  = {Or{\'u}s, Rom{\'a}n},
  journal = {Annals of Physics},
  volume  = {349},
  pages   = {117--158},
  year    = {2014},
  doi     = {10.1016/j.aop.2014.06.013}
}

@article{schollwoeck2011dmrg,
  title   = {The density-matrix renormalization group in the age of matrix product states},
  author  = {Schollw{\"o}ck, Ulrich},
  journal = {Annals of Physics},
  volume  = {326},
  number  = {1},
  pages   = {96--192},
  year    = {2011},
  doi     = {10.1016/j.aop.2010.09.012}
}

@article{hubig2017generic,
  title   = {Generic construction of efficient matrix product operators},
  author  = {Hubig, Claudius and McCulloch, Ian P. and Schollw{\"o}ck, Ulrich},
  journal = {Physical Review B},
  volume  = {95},
  number  = {3},
  pages   = {035129},
  year    = {2017},
  doi     = {10.1103/PhysRevB.95.035129}
}

@article{paeckel2017automated,
  title   = {Automated construction of {U(1)}-invariant matrix-product operators from graph representations},
  author  = {Paeckel, Sebastian and K{\"o}hler, Thomas and Manmana, Salvatore R.},
  journal = {SciPost Physics},
  volume  = {3},
  number  = {5},
  pages   = {035},
  year    = {2017},
  doi     = {10.21468/SciPostPhys.3.5.035}
}

@article{ren2020automatic,
  title   = {A general automatic method for optimal construction of matrix product operators using bipartite graph theory},
  author  = {Ren, Jiajun and Li, Weitang and Jiang, Tong and Shuai, Zhigang},
  journal = {The Journal of Chemical Physics},
  volume  = {153},
  number  = {8},
  pages   = {084118},
  year    = {2020},
  doi     = {10.1063/5.0018149}
}

@article{pirvu2010matrix,
  title   = {Matrix product operator representations},
  author  = {Pirvu, Bogdan and Murg, Valentin and Cirac, J Ignacio and Verstraete, Frank},
  journal = {New Journal of Physics},
  volume  = {12},
  number  = {2},
  pages   = {025012},
  year    = {2010}
}

@article{lucas2014ising,
  title={Ising formulations of many NP problems},
  author={Lucas, Andrew},
  journal={Frontiers in physics},
  volume={2},
  pages={74887},
  year={2014},
  publisher={Frontiers}
}

@article{haegeman2017diagonalizing,
  title={Diagonalizing transfer matrices and matrix product operators: A medley of exact and computational methods},
  author={Haegeman, Jutho and Verstraete, Frank},
  journal={Annual Review of Condensed Matter Physics},
  volume={8},
  number={1},
  pages={355--406},
  year={2017},
  publisher={Annual Reviews}
}

\clearpage
\section*{Current executable software version}

\begin{table}[H]
\centering
\begin{tabularx}{\linewidth}{|l|>{\raggedright\arraybackslash}X|}
\hline
\textbf{Executable metadata} & \textbf{Value} \\
\hline
Software name & \texttt{autotn} \\
\hline
Version & 1.0.0 \\
\hline
Programming language & Python \\
\hline
Main dependency & NumPy \\
\hline
Test command & \texttt{python -m pytest} \\
\hline
Repository & \href{https://github.com/aitormorais/tngenerator/tree/v1.0.0}{GitHub tag v1.0.0}. \\
\hline
\end{tabularx}
\caption{Executable software metadata.}
\end{table}

\end{document}